\documentclass[prd,aps,a4paper,nofootinbib,twocolumn]{revtex4}   

\newif\ifusesec
\usesectrue  
   
\usepackage{mathrsfs}
\usepackage{amsmath,amsfonts,amssymb}

\newcommand{\beq}{\begin{equation}}
\newcommand{\eeq}{\end{equation}}

\def\rightcontract{\mathop{\hbox{\vrule width0.5pt height6pt%
  \vrule height0.5pt width6pt}}}

\begin{document}

\title{Black hole geodesic parallel transport and the Marck recipe for isolating cumulative precession effects }

\author{Donato \surname{Bini}$^{1,2,3}$}
\author{Andrea \surname{Geralico}$^1$}
\author{Robert T. \surname{Jantzen}$^{3,4}$}

\affiliation{
$^1$Istituto per le Applicazioni del Calcolo ``M. Picone'', CNR, I--00185 Rome, Italy\\
$^2$INFN Sezione di Napoli, Complesso Universitario di Monte S. Angelo,
Via Cintia Edificio 6, 80126 Naples, Italy\\
$^3$ International Center for Relativistic Astrophysics Network, I--65122 Pescara, Italy\\
$^4$Department of Mathematics and Statistics, Villanova University, Villanova, PA 19085, USA
}

\date{\today}
\begin{abstract}
The Wigner rotations arising from the combination of boosts along two different directions are rederived from a relative boost point of view and applied to gyroscope spin precession along timelike geodesics in a Kerr spacetime, clarifying the geometrical properties of Marck's recipe for describing parallel transport along such world lines expressed in terms of the constants of the motion. His final angular velocity isolates the cumulative spin precession angular velocity independent of the spacetime tilting required to keep the spin 4-vector orthogonal to the gyro 4-velocity.
As an explicit example the cumulative precession effects are computed for a test gyroscope moving along both bound and unbound equatorial plane geodesic orbits.
\end{abstract}

\maketitle

\section{Introduction}

Given any two forward pointing timelike unit vectors 
$u_1$ and  $u_2$ in the tangent space to a Lorentzian spacetime (signature $+2$), which may be interpreted as the 4-velocities of a pair of test observers at that event, there is a unique active Lorentz transformation $B(u_2,u_1)$ which takes one ($u_1$) into the other ($u_2$), termed a relative observer boost, acting only in the plane of the two vectors as a hyperbolic rotation. This boost is most conveniently parametrized by the hyperbolic rotation angle, often called the rapidity $\beta={\rm arccosh}(-u_1\cdot u_2 )$. For boosting between three such successive 4-velocities in the same plane, the rapidity parameters are additive, but when their 4-velocities are not coplanar,  two successive relative observer boosts are equivalent to a single such boost followed by a rotation, called the Wigner rotation \cite{Wigner:1939cj}, due to the fact that the boost generators of the Lorentz group do not form a closed Lie subalgebra, but their commutators lead to rotation generators. All of the calculations in this case only involve special relativity. Furthermore, to combine boosts with the Lorentz group multiplication law, they should all be referred to a common time direction say $u$, i.e., $B(u_2,u) B(u_1,u) X$ when applied to some spacetime vector $X$ \cite{Ferraro:1999eu,Bakke:2015yia,O'Donnell:2011am}.

The Wigner rotation is intimately connected with the Thomas precession effect in special relativity, most notably studied for a classical spinning electron in a circular orbit \cite{Thomas:1926dy,Thomas:1927yu,furry,shelupsky,mackenzie,fischer,goedecke,muller,Rhodes:2003id,Han:1987gj}. The Thomas precession is a dynamical expression of the instantaneous Wigner rotation effect, in the context of the succession of boosts from the laboratory frame to the particle rest frame along the changing direction of the particle trajectory. This takes a slightly different form in the general relativistic analysis of spin precession of a test gyroscope in a given curved spacetime through Fermi-Walker transport of the spin 4-vector, but the Wigner rotation and a generalized Thomas precession remains important. Along geodesic world lines, Fermi-Walker transport reduces to parallel transport.

In the study of the precession of the parallel transported spin vector of a test gyroscope moving along a geodesic orbit in a rotating Kerr black hole spacetime, the key to geodesic motion and parallel transport is the Carter orthonormal frame \cite{Carter:1968ks} in Boyer-Lindquist coordinates $\{t,r,\theta,\phi\}$
which together with the Killing vectors $\partial_t$ and $\partial_\phi$, symmetric Killing 2-tensor $K_{\alpha\beta}$ and Killing-Yano 2-form $f_{\alpha\beta}$ leads to the separability of the geodesic equations and the solution of the equations of parallel transport along geodesics, modulo first order differential equations expressible entirely in terms of the constants of the motion. The Carter orthonormal frame is boosted along the azimuthal direction associated with the rotational Killing vector field compared to the usual static observer frame, the latter observers having world lines which are the time lines of the Boyer-Lindquist coordinate system. The Carter frame corotates with the black hole and is well-defined everywhere outside its outer horizon.

The Carter frame differs only by a boost from the more familiar slicing and threading spherical frames which are obtained from the Boyer-Lindquist coordinate system. The slicing frame or zero-angular-momentum-observer (ZAMO) frame is obtained by normalizing the orthogonal spatial coordinate frame to a spherical orthonormal triad and completing it by adding the unit normal $n$ to the time coordinate hypersurfaces. The threading frame adapted to the so called static observers following the time coordinate lines (when timelike, having 4-velocity $m$) differs only by a boost in the 2-plane of the two Killing vector fields tangent to the cylinders of the $t$-$\phi$ coordinates, namely a boost along the azimuthal direction. Similarly the Carter frame also only differs by a boost in that same 2-plane. Each of these boosts just reflects how different observers in relative rotation around the azimuthal direction see the same orthonormal triad of vector fields rigidly attached to the coordinate system, which in turn is rigidly connected to ``the distant stars" (spatial infinity) in the sense that the threading observers see the incoming light rays from those distant stars forming a time-independent pattern on the celestial sphere. Since the spin vector along any timelike world line remains in the local rest space of that world line, one can also boost the spherical triad to that local rest space in order to measure the rotation of the spin with respect to the triad of vectors as seen by an observer following that world line. In a sense this subtracts the spherical aberration of the incoming light rays from spatial infinity on the local celestial sphere.  Whether one boosts one of the other boosted spherical frames or performs a direct boost from the static frame to the local rest space of a gyro, the results differ only by orientation induced Wigner rotations reflecting the relative tilting of the local rest spaces and do not add to any accumulating angle of precession.

The local rest space of the gyro's geodesic world line is related to the Carter frame by two successive relative observer boosts at right angles, first along the radial direction which preserves the radial alignment of the electric and magnetic parts of the Killing-Yano 2-form while eliminating the radial relative velocity of the gyro, and then along an orthogonal angular direction to reach the local rest space of the geodesic by eliminating its angular velocity. Thus a sequence of three successive relative observer boosts, each at right angles to each other, take the static observer frame to a frame aligned with the geodesic 4-velocity. This geodesic frame is rotated with respect to a direct relative observer boost of the static observer axes to the gyro local rest space, which are an important comparison reference frame with respect to which the spin angular velocity precession has the stellar aberration effect subtracted away.   Each pair of successive boosts leads to a Wigner rotation with respect to the equivalent direct boost. The result of the three successive boosts is then each rotated with respect to the direct boost. One can evaluate each of these rotations. 

Starting from the Carter frame, Marck's construction \cite{marck1,marck2} of a parallely propagated frame along a timelike geodesic with 4-velocity $U$ utilizes the electric part $f\rightcontract U$ of the Killing-Yano 2-form as seen by the test observer following the gyro. This vector is parallel transported along the world line and defines the parallel transported normal direction to the 2-plane of the parallel transport rotation within the local rest space $LRS_U$ of the gyro \cite{Jantzen:1992rg}. The rest of his construction uses a sequence of boosts along the radial and angular directions to determine a natural pair of orthonormal vectors in this plane to express the angular velocity of the parallel transport rotation within it. Ultimately it is the comparison of this angular velocity of rotation with the local static Cartesian frame associated with the spherical frame which allows one to extract the spin precession with respect to the distant stars.

We begin by rederiving the Wigner rotation associated with two successive boosts and then use it to analyze the geometrical meaning of the Marck construction of a parallely propagated frame along general timelike geodesics in a Kerr black hole spacetime.

We use the signature $-+++$ and Greek and Latin index conventions $\alpha,\beta,\gamma=0,1,2,3$, $i,j,k=1,2,3$.

\section{A relative observer boost}

Consider two 4-velocities in the same tangent space, $u$ and $U$. The orthogonal decomposition of $U$ with respect to $u$ and its local rest space $LRS_u$ defines a relative velocity $\nu(U,u)$ and unit direction vector $\hat\nu(U,u)=\nu(U,u)/||\nu(U,u)||$
\begin{eqnarray}
\label{U_split}
  U&=&\gamma(U,u) \left[ u + \nu(U,u) \right]\nonumber\\
   &=& \cosh \alpha\, u + \sinh \alpha \, \hat\nu(U,u) \,,
\end{eqnarray}
where the associated gamma factor has the usual expression
$\gamma(U,u) = (1-||\nu(U,u)||^2)^{-1/2} =\cosh \alpha $ defining the rapidity $\alpha\ge0$,
in terms of which the relative speed is $||\nu(U,u)||=\tanh \alpha $.
An active relative observer boost $B(U,u)$ of the tangent space takes $u$ onto $U=B(U,u)u$ and acts as the identity orthogonal to the plane of $u$ and $U$, mapping the local rest space $LRS_u$ onto $LRS_U$ 
 \cite{Jantzen:1992rg,Bini:1995hx,Felice:2010rpa}.

Let $P(u)=Id+u\otimes u^\flat$, $P(U)=Id+U\otimes U^\flat$ be the mixed tensors representing projections onto these respective local rest spaces, with $Id$ denoting the identity tensor.
Then restricting this boost to a map $B_{\rm (lrs)}(U,u)=P(U)B(U,u)P(u)$
from $LRS_u$ onto $LRS_U$ 
(see Eq.~4.22 of Ref.~\cite{Jantzen:1992rg} for additional details), one finds
\beq
\label{B_lrs}
B_{\rm (lrs)}(U,u)   S= S+\frac{(U\cdot S)}{\gamma(U,u)+1} (U+u) \,,
\eeq
for any vector $S$ belonging to the local rest space of $u$ 
($S\cdot u=0=U\cdot [B_{\rm (lrs)}(U,u)   S]$). 
Adjacent $1\choose 1$ tensors are understood to be contracted on their adjacent indices reflecting the composition of the corresponding linear maps of the tangent space.
To express the full boost in this notation,
let
\beq
X=X^{(\Vert)} u + X^{(\perp)}
\eeq
with
\beq
X^{(\Vert)} =-u\cdot X\,,\quad
X^{(\perp)} = X+(u\cdot X)u=P(u)X
\eeq
be a generic spacetime vector orthogonally decomposed with respect to $u$.
Then one finds 
\begin{eqnarray}
\label{Boost_expl}
B(U,u)  X &=& B(U,u) (X^{(\Vert)} u + X^{(\perp)})\,,
\nonumber\\
 &=& X^{(\Vert)} U + X^{(\perp)} + \frac{X^{(\perp)}\cdot U}{\gamma(U,u)+1} (u+U)
\nonumber\\
&=&
 \left[ Id -\frac{1}{\gamma+1} [(2\gamma+1) U -u]\otimes u^\flat \right. \nonumber\\
&& \left.+ \frac{1}{\gamma+1} (U+u)\otimes U^\flat\right]  X
\end{eqnarray}
with the shorthand abbreviation $\gamma=\gamma(U,u)$.

Replacing $U$ by Eq.~\eqref{U_split} and then using the identity
$Id=P(u)-u\otimes u^\flat$,
one finds
\begin{eqnarray}
\label{Boost_expl}
 B(U,u) 
&=&  Id -(\gamma-1)[ u\otimes u^\flat -\hat \nu \otimes \hat \nu^\flat ] \nonumber\\
&& -\gamma \nu [\hat \nu \otimes u^\flat -u \otimes \hat \nu ^\flat] 
\nonumber\\
&=&  
B_{{\rm (lrs)}u}(U,u)
-\gamma \nu(U,u)\otimes u^\flat\nonumber\\
&&  +\gamma u \otimes \nu(U,u)^\flat
-\gamma u\otimes u^\flat\,,
\end{eqnarray}
where  
\begin{eqnarray}
B_{{\rm (lrs)}u}(U,u)
&=& P(u)  B(U,u)  P(u)
\nonumber\\
&=& P(u)+\frac{\gamma^2}{\gamma+1}\nu(U,u)\otimes \nu(U,u)^\flat
\nonumber\\
&=& P(u)+{(\gamma-1)}\hat\nu(U,u)\otimes \hat\nu(U,u)^\flat\,.
\end{eqnarray}
Its inverse map is
\begin{eqnarray}
[B_{{\rm (lrs)}u}(U,u)]^{-1}
&=& B_{{\rm (lrs)}u}(u,U)
\nonumber\\
&=& P(u)  B(u,U) P(u)
\nonumber\\
&=&P(u)-{(\gamma-1)}\hat\nu(U,u)\otimes \hat\nu(U,u)^\flat\,.\nonumber\\
\end{eqnarray}

Let $e(u)_a$ be an orthonormal spatial triad adapted to $u=e(u)_0$ and let $\omega(u)^\alpha$ be the dual frame with  $\omega(u)^0=-u^\flat$.
The frame components of $B(U,u)$ with respect to this frame
\beq
 B(U,u)^\alpha{}_\beta = \omega^\alpha(B(U,u)  e_\beta)\,,
\eeq
are then explicitly
\begin{eqnarray}
  B^0{}_0 &=& \gamma\,,\quad
  B^0{}_a = \gamma \nu(U,u)_a\,,\quad
  B^a{}_0 = \gamma \nu(U,u)^a\,,\nonumber\\
  B^a{}_b &=& P(u)^a{}_b +{(\gamma-1)} \hat\nu(U,u)^a\hat\nu(U,u)_b\,.
\end{eqnarray}

Defining the generator  $K(u)_i=e_i(u) \otimes u^\flat-u \otimes e_i^\flat$ of boosts in the direction $e(u)_i$, one has the following representation for the generator of a boost in the direction $\hat\nu(U,u)$
\beq
K(u)_i  \hat\nu(U,u)^i =  \hat \nu \otimes u^\flat -u \otimes \hat \nu ^\flat\,.
\eeq
Then
\begin{eqnarray}
(K(u)_i  \hat\nu(U,u)^i)^2
&=& (K(u)_i  \hat\nu(U,u)^i)  (K(u)_j  \hat\nu(U,u)^j)\nonumber\\
&=& -u\otimes u^\flat +\hat \nu \otimes \hat \nu ^\flat \,,
\end{eqnarray}
and our previous expression \eqref{Boost_expl} becomes
\begin{eqnarray}
B(U,u) &=&Id +(\gamma-1)(\hat \nu \cdot K(u))^2 +\gamma \nu (\hat \nu \cdot K(u)) \nonumber\\
&=&
Id +(\cosh \alpha-1)(K(u)_i  \hat\nu(U,u)^i)^2\nonumber\\
&& +\sinh \alpha\, (K(u)_i  \hat\nu(U,u)^i)
\nonumber\\
&=& e^{\alpha\, K(u)_i  \hat\nu(U,u)^i}\,.
\end{eqnarray}
This corresponds directly to a matrix relation (see Eq.~\eqref{rotationboostmatrices} of the Appendix A below)  when expressed in an adapted orthonormal frame.
For instance, in the simplest case of $\hat\nu^a=(\cos\delta,0,\sin\delta)$ we have

\begin{widetext}

\beq
(B(U,u)^\alpha{}_\beta )=
\begin{pmatrix}
\cosh\alpha& \sinh\alpha\cos\delta& 0 &\sinh\alpha\sin\delta\cr
\sinh\alpha\cos\delta& \cosh\alpha\cos^2\delta+\sin^2\delta& 0& \displaystyle\frac{\sinh^2\alpha\cos\delta\sin\delta}{1+\cosh\alpha}\cr
0& 0& 1& 0\cr
\sinh\alpha\sin\delta& \displaystyle\frac{\sinh^2\alpha\cos\delta\sin\delta}{1+\cosh\alpha}& 0& \cosh\alpha\sin^2\delta+\cos^2\delta\cr
\end{pmatrix}
\,.
\eeq

\end{widetext}

\subsection{Composition of velocities}

Consider expressing a single 4-velocity $U$ in terms of two distinct observer 4-velocities $u$ and $u'$
\beq
U=\underbrace{\gamma(U,u)[u+\nu(U,u)]}_{U \, {\rm vs}\, u}=\underbrace{\gamma(U,u')[u'+\nu(U,u')]}_{U \, {\rm vs}\, u'}
\eeq
with
\begin{eqnarray}
u&=&\underbrace{\gamma(u,u')[u'+\nu(u,u')]}_{u \, {\rm vs}\, u'}\,,\nonumber\\ 
u'&=&\underbrace{\gamma(u,u')[u+\nu(u',u)]}_{u' \, {\rm vs}\, u}\,.
\end{eqnarray}
A straightforward calculation shows that the gamma factors are related by
\beq
\frac{\gamma(U,u')}{\gamma(U,u)}=\gamma(u,u')[1-\nu(U,u)\cdot \nu(u',u)]
\eeq
and the ``addition of velocity" formula is 
\beq
P(u',u)  \left[\frac{\nu(U,u)-\nu(u',u)}{1-\nu(U,u)\cdot \nu(u',u)}  \right]=\gamma(u,u') \nu(U,u')\,.
\eeq
When considering the action of the boost map $B(U,u) P(u)$ on the relative velocity of $U$ with respect to $u$
we find 
\begin{widetext}
\begin{eqnarray}
[B(U,u) P(u)]\nu(U,u)&=& [P(u) + \frac{\gamma(U,u)} {\gamma(U,u)+1} (u+U) \otimes \nu(U,u)] \nu(U,u)\nonumber\\
&=& \nu(U,u) + \frac{\gamma(U,u) ||\nu(U,u)||^2} {\gamma(U,u)+1} (u+U)\nonumber\\
&=& \nu(U,u) + \frac{\gamma(U,u) -1} {\gamma(U,u)} (u+U)
=-\nu(u,U)\,.
\end{eqnarray}
\end{widetext}
Equivalently, since $P(u)\nu(U,u)=\nu(U,u)$, then 
$B(U,u)\nu(U,u)=-\nu(u,U)$,
as in Eq.~(4.5) of \cite{Jantzen:1992rg}.

\section{Combination of boosts}

Boosts along a fixed spatial direction form a subgroup of the Lorentz group but the subset of boosts do not form a subgroup: the product of two boosts along distinct spatial directions is no longer a boost along any direction but instead the product of a single boost with a rotation, the Wigner rotation. To discuss them, it is essential to introduce the matrix representation of boosts and rotations to divorce them from the various orthonormal frames in which our tensor representations can be expressed. The notation $B(u_2,u_1)$ facilitates expressing boosts from one local rest space to the next  but is simple only when expressed in terms of an orthonormal frame adapted to the initial local rest space $LRS_{u_1}$ so when composed with a second boost, that matrix representation of the successive boost is complicated.
 
Let $u=e_0$ denote the future-pointing timelike vector (4-velocity) in an orthonormal frame $\{e_\alpha\}$ with dual frame $\{\omega^\alpha\}$, and define
\beq
  {\mathcal B}({\pmb \nu})^\alpha{}_\beta = \omega^\alpha(B(U,u) e_\beta)
\,,
\eeq
or
\beq  
B(U,u) e_\beta = e_\alpha{\mathcal B}({\pmb \nu})^\alpha{}_\beta \,,
\eeq
where   ${\pmb\nu}=\langle \nu^1,\nu^2,\nu^3 \rangle $ is an ordinary 3-tuple vector of the spatial orthonormal components of the relative velocity $\nu$ of a 4-velocity $U$ with respect to $u$. See Appendix A for explicit formulas.

Consider the combinations of two boosts in different directions.
To this end let  $u_1$, $u_2$ and $u_3$ be three 4-velocities, with the last one determining the final local rest space of the combination of Lorentz boosts associated with $u_1$ and $u_2$ with respect to the common time direction $u$
\begin{eqnarray}
u_3 &=& B(u_1,u)u_2\\
    &=& B(u_1,u)B(u_2,u)u
\end{eqnarray}
and define the relative gamma factors and relative velocities by 
\beq
u_i = \gamma_i (u+ {\pmb\nu}_i)\,,
\eeq
where it is convenient to use the abbreviated notation
$(\gamma_i,{\pmb \nu}_i) = (\gamma(u_i,u), \nu (u_i,u))$.

Consider two successive boosts of a vector $X=X^a e_a$ ($a=1,2,3$) belonging to $LRS_u$, namely
\beq
  B(u_1,u) B(u_2,u)  X  
= {\mathcal B}({\pmb \nu_1})^\alpha{}_\gamma  {\mathcal B}({\pmb \nu_2})^\gamma{}_\beta X^\beta\,.
\eeq
The  first boost takes $X\in LRS_u$ to $LRS_{u_2}$ (while taking $u$ to $u_2$) and the next boost $B(u_1,u)$ takes that vector to $LRS_{u_3}$ (while taking $u_2$ to $u_3$ which determines the final local rest space $LRS_{u_3}$ of the combined boosts). 
One can also directly boost $X$ from $LRS_u$ to $LRS_{u_3}$, namely $B(u_3,u)X$.
These two vectors both belong to the final local rest space but differ by a Wigner rotation defined by
\beq
B(u_1,u) B(u_2,u) X
  = B(u_3,u) R^{\rm (W)}(u_1,u_2;u) X\,,
\eeq
with solution
\begin{eqnarray}
 R^{\rm (W)}(u_1,u_2;u)  X
&=&B(u_3,u)^{-1}  B(u_1,u) B(u_2,u) X \nonumber\\
&=&B(u,u_3)  B(u_1,u) B(u_2,u) X  \,.
\end{eqnarray}
Explicit calculation of $u_3$ using Eq.~\eqref{Boost_expl} 
(or by multiplication of the corresponding matrices using a computer algebra system)
shows that
\beq
  \gamma_3 = \gamma_1\gamma_2 (1+{\pmb\nu}_1\cdot {\pmb \nu}_2)
\eeq
and  
\beq
 {\pmb\nu}_3 = \frac{1}{1+{\pmb \nu}_1\cdot {\pmb \nu}_2} \left[ \left(1+\frac{\gamma_1 {\pmb\nu}_1\cdot {\pmb\nu}_2}{1+\gamma_1} \right) {\pmb\nu}_1+\frac{1}{\gamma_1} {\pmb\nu}_2\right] \,,
\eeq
namely, ${\pmb\nu}_3$ belongs to the 2-plane spanned by  ${\pmb\nu}_1$ and ${\pmb\nu}_2$.
Without any loss of generality one can align the axis $e(u)_1$ with  ${\pmb \nu}_1$ and the axis $e(u)_3$ orthogonal to the 2-plane spanned by  ${\pmb \nu}_1$ and ${\pmb \nu}_2$, that is
\begin{eqnarray}
{\pmb \nu}_1&=& \tanh (\alpha_1)e(u)_1
\,,\nonumber\\
{\pmb \nu}_2&=& \tanh (\alpha_2) [\cos \beta_2\, e(u)_1+\sin \beta_2\, e(u)_2]
\,.
\end{eqnarray}
In this case we find
\begin{eqnarray}
{\pmb \nu}_1 \cdot {\pmb \nu}_2 &=& \tanh\alpha_1 \tanh\alpha_2 \cos \beta_1\,,\nonumber\\
{\pmb \nu}_3 &=& \tanh \alpha_3 \,[\cos \beta_3\, e(u)_1+\sin \beta_3\, e(u)_2]\,,
\end{eqnarray}
with
\beq
\cosh \alpha_3 = \cosh \alpha_1 \cosh \alpha_1  (1+\tanh\alpha_1 \tanh\alpha_2 \cos \beta_2)
\eeq
and
\beq
\label{beta_3_eq}
\tan \beta_3 = \frac{\nu_2\sin \beta_2}{\gamma_1 (\nu_1+\nu_2\cos \beta_2)} \,.
\eeq
For boosts in parallel directions, say $e(u)_1$, ${\pmb \nu}_1=\tanh  \alpha_1  e(u)_1$, ${\pmb \nu}_2=\tanh \alpha_2  e(u)_1$ and  ${\pmb \nu}_3=\tanh  \alpha_3  e(u)_1$ (i.e., $\beta_2=0$) the rapidity is simply additive,
\beq
\alpha_3  = \alpha_1+\alpha_2  \,,
\eeq
reflecting the fact that boosts along a fixed direction form a subgroup,
and the rotation reduces to the identity, while
for the complementary case of boosts in orthogonal directions so that ${\pmb \nu}_1\cdot {\pmb \nu}_2=0$ ($\beta_2=\pi/2$),  it turns out that the rotation angle is maximized, and these formulas reduce to
\beq
  \gamma_3 = \gamma_1\gamma_2\,,\quad
{\pmb \nu}_3 =  {\pmb \nu}_1+\frac{1}{\gamma_1} {\pmb \nu}_2 \,,
\eeq
since the first applied boost velocity ${\pmb \nu}_2$ must be adjusted to the proper time of the second applied boost (ordered right to left).
The Wigner rotation  results 
\beq
\label{wigner_36}
(R^{\rm (W)}(u_1,u_2;u)^\alpha{}_\beta )= 
\begin{pmatrix}
1 & 0& 0&  0 \cr
0 & \cos \theta & \sin \theta  & 0 \cr
0& -\sin \theta  & \cos \theta  & 0 \cr
0 & 0& 0 &  1 \cr
\end{pmatrix}
\,,
\eeq
with
\begin{widetext}
\begin{eqnarray}
\label{final_rel}
\sin\theta &=& \left(1+\frac{\gamma_1+\gamma_2}{1+\gamma_2\gamma_1(1+ \nu_1\nu_2\cos\beta_2)}  \right)
\frac{\gamma_1 \gamma_2 \nu_1 \nu_2}{(1+\gamma_1)(1+\gamma_2)} \sin \beta_2\,,\\
\label{final_rel2}
\cos\theta 
&=& 1- \frac{(1-\gamma_1)(1-\gamma_2)}{1+\gamma_2\gamma_1(1+ \nu_1\nu_2\cos\beta_2)}\sin^2 \beta_2\,.
\end{eqnarray}
\end{widetext}

For the case of successive boosts along orthogonal directions $\beta_2=\pi/2$ the above relation reduces to
\beq
\cos \theta  = \frac{ \gamma_1+\gamma_2 }{ 1+\gamma_1\gamma_2 }
\,,\qquad
\sin \theta  = \frac{\gamma_1 \nu_1\gamma_2 \nu_2}{ 1+\gamma_1\gamma_2 }\,.
\eeq

Eqs.~\eqref{final_rel} and \eqref{final_rel2} have extreme values at $\nu_1= \nu_2$ and $\beta_2=\pm \pi/2$ 
\beq
 \sin\theta_{\rm (ext)} 
=\pm \frac{\gamma_1^2-1}{\gamma_1^2+1}\, 
=\pm \frac{\sinh^2\alpha_1}{1+\cosh^2\alpha_1}  \,,
\eeq 
or
\beq
 \cos\theta_{\rm (ext)} 
=\pm \frac{2\gamma_1}{1+\gamma_1^2}\, 
=\pm \frac{2\cosh\alpha_1}{1+\cosh^2\alpha_1}  \,,
\eeq 
which is confined to the interval $(-1,1)$ and $-\frac{\pi}{2}<\theta_{\rm (ext)}<\frac{\pi}{2}$.

For concreteness consider the special examples of two successive boosts along a pair of orthogonal directions aligned with the frame vectors $\{e(u)_i\}$. 
We have formally the same expression
\beq
{\mathcal B}({\pmb \nu_1})^\alpha{}_\gamma  {\mathcal B}({\pmb \nu_2})^\gamma{}_\beta=
{\mathcal B}({\pmb \nu_3})^\alpha{}_\gamma  R^{\rm (W)}(u_1,u_2;u)^\gamma{}_\beta\,,
\eeq
where ${\mathcal B}({\pmb \nu_3})$ and $R^{\rm (W)}(u_1,u_2;u)$ will be specified case by case.

\begin{widetext}

\begin{enumerate}
  
\item ${\pmb \nu}_1=\tanh\alpha_1\, e(u)_1$, ${\pmb \nu}_2=\tanh\alpha_2\, e(u)_2$, so that  ${\pmb \nu}_3=\tanh\alpha_1\, e(u)_1+(\tanh\alpha_2/\cosh\alpha_1)\,e(u)_2$.
We find
\beq
({\mathcal B}({\pmb \nu_3})^\alpha{}_\gamma )= 
\begin{pmatrix}
\cosh\alpha_1\cosh\alpha_2 & \sinh\alpha_1\cosh\alpha_2& \sinh\alpha_2& 0 \cr
\sinh\alpha_1\cosh\alpha_2 & 1+\displaystyle\frac{\sinh^2\alpha_1\cosh^2\alpha_2}{1+\cosh\alpha_1\cosh\alpha_2}& \displaystyle\frac{\cosh\alpha_2\sinh\alpha_1\sinh\alpha_2}{1+\cosh\alpha_1\cosh\alpha_2}& 0 \cr
\sinh\alpha_2 & \displaystyle\frac{\cosh\alpha_2\sinh\alpha_1\sinh\alpha_2}{1+\cosh\alpha_1\cosh\alpha_2}& \displaystyle\frac{\cosh\alpha_2(\cosh\alpha_1+\cosh\alpha_2)}{1+\cosh\alpha_1\cosh\alpha_2}& 0 \cr
0 &0 &0 &1  \cr
\end{pmatrix}
\,,
\eeq
and
\beq
(R^{\rm (W)}(u_1,u_2;u)^\gamma{}_\beta )= 
\begin{pmatrix}
1 &0 &0 &0  \cr
0 &\cos\theta & \sin\theta&0 \cr
0 &-\sin\theta &\cos\theta &0  \cr
0 &0 &0 &1  \cr
\end{pmatrix}
\,,
\eeq
as in Eq.~\eqref{wigner_36}, with 
\beq
\label{sincosth}
\cos\theta=\frac{\cosh\alpha_1+\cosh\alpha_2}{1+\cosh\alpha_1\cosh\alpha_2}\,,\qquad
\sin\theta=\frac{\sinh\alpha_1\sinh\alpha_2}{1+\cosh\alpha_1\cosh\alpha_2}\,.
\eeq

\item ${\pmb \nu}_1=\tanh (\alpha_1) e(u)_1$, ${\pmb \nu}_2=\tanh (\alpha_2) e(u)_3$, so that  ${\pmb \nu}_3=\tanh\alpha_1\, e(u)_1+(\tanh\alpha_2/\cosh\alpha_1)\,e(u)_3$.
We find
\beq
{\mathcal B}({\pmb \nu_1})^\alpha{}_\gamma  {\mathcal B}({\pmb \nu_2})^\gamma{}_\beta=
{\mathcal B}({\pmb \nu_3})^\alpha{}_\gamma  R^{\rm (W)}(u_1,u_2;u)^\gamma{}_\beta\,,
\eeq
with
\beq\label{Bnu3}
({\mathcal B}({\pmb \nu_3})^\alpha{}_\gamma )= 
\begin{pmatrix}
\cosh\alpha_1\cosh\alpha_2 & \sinh\alpha_1\cosh\alpha_2& 0& \sinh\alpha_2 \cr
\sinh\alpha_1\cosh\alpha_2 & 1+\displaystyle\frac{\sinh^2\alpha_1\cosh^2\alpha_2}{1+\cosh\alpha_1\cosh\alpha_2}& 0& \displaystyle\frac{\cosh\alpha_2\sinh\alpha_1\sinh\alpha_2}{1+\cosh\alpha_1\cosh\alpha_2} \cr
0 &0 &1 &0  \cr
\sinh\alpha_2 & \displaystyle\frac{\cosh\alpha_2\sinh\alpha_1\sinh\alpha_2}{1+\cosh\alpha_1\cosh\alpha_2}& 0& \displaystyle\frac{\cosh\alpha_2(\cosh\alpha_1+\cosh\alpha_2)}{1+\cosh\alpha_1\cosh\alpha_2} \cr
\end{pmatrix}
\,,
\eeq
and
\beq
(R^{\rm (W)}(u_1,u_2;u)^\gamma{}_\beta )= 
\begin{pmatrix}
1 &0 &0 &0  \cr
0 &\cos\theta & 0& \sin\theta\cr
0 &0 &1 &0  \cr
0 &-\sin\theta &0& \cos\theta  \cr
\end{pmatrix}
\,,~
\eeq
with $\cos\theta$ and $\sin\theta$ still given by Eq.~\eqref{sincosth}.

\item ${\pmb \nu}_1=\tanh (\alpha_1) e(u)_2$, ${\pmb \nu}_2=\tanh (\alpha_2) e(u)_3$, so that  ${\pmb \nu}_3=\tanh\alpha_1\, e(u)_2+(\tanh\alpha_2/\cosh\alpha_1)\,e(u)_3$.
We find
\beq
({\mathcal B}({\pmb \nu_3})^\alpha{}_\gamma )= 
\begin{pmatrix}
\cosh\alpha_1\cosh\alpha_2 & \sinh\alpha_1\cosh\alpha_2& 0& \sinh\alpha_2 \cr
0 &1 &0 &0  \cr
\sinh\alpha_1\cosh\alpha_2 & 0& 1+\displaystyle\frac{\sinh^2\alpha_1\cosh^2\alpha_2}{1+\cosh\alpha_1\cosh\alpha_2}& \displaystyle\frac{\cosh\alpha_2\sinh\alpha_1\sinh\alpha_2}{1+\cosh\alpha_1\cosh\alpha_2} \cr
\sinh\alpha_2 & 0& \displaystyle\frac{\cosh\alpha_2\sinh\alpha_1\sinh\alpha_2}{1+\cosh\alpha_1\cosh\alpha_2}& \displaystyle\frac{\cosh\alpha_2(\cosh\alpha_1+\cosh\alpha_2)}{1+\cosh\alpha_1\cosh\alpha_2} \cr
\end{pmatrix}
\,,
\eeq
and
\beq
(R^{\rm (W)}(u_1,u_2;u)^\gamma{}_\beta )= 
\begin{pmatrix}
1 &0 &0 &0  \cr
0 &1 &0 &0  \cr
0 &0 &\cos\theta & \sin\theta \cr
0 &0 & -\sin\theta & \cos\theta  \cr
\end{pmatrix}
\,,
\eeq
with $\cos\theta$ and $\sin\theta$ still given by Eq.~\eqref{sincosth}.

\end{enumerate}  

\end{widetext}

\section{Special observers and adapted frames in the Kerr  spacetimes}

Let us consider the Kerr spacetime with metric written in the Boyer-Lindquist coordinate system $(t,r,\theta,\phi)$ \cite{Misner:1974qy},
\begin{eqnarray}
\label{K1}
ds^2&=&g_{\alpha\beta}dx^\alpha dx^\beta\nonumber\\
&=&-dt^2+\frac{\Sigma}{\Delta}dr^2+\Sigma\, d\theta^2 +(r^2+a^2)\sin^2\theta\, d\phi^2\nonumber\\
&& +\frac{2Mr}{\Sigma}(dt-a\sin^2\theta\, d\phi)^2\,,
\end{eqnarray}
where $M$ and $a$ are the mass and the specific angular momentum of the  source, respectively, and
\beq\label{K2}
\Sigma=r^2+a^2\cos^2\theta\,,\qquad \Delta=r^2-2Mr+a^2\,.
\eeq
The inner and outer horizons are located at $r_\pm=M\pm\sqrt{M^2-a^2}$.

In this spacetime there exist at  least three families of fiducial/special observers who play a role from either a geometrical point of view or a physical point of view. They are zero-angular-momentum observers (ZAMOs, with 4-velocity $u=n$), static observers (with 4-velocity $u=m$) and Carter observers (with 4-velocity $u=u_{\rm (car)}$).

The ZAMOs have world lines orthogonal to the Boyer-Lindquist $t=$constant hypersurfaces, the static observers have world lines aligned with 
Boyer-Lindquist temporal lines and the Carter observers have 4-velocity belonging to the intersection of the two 2-planes: the one spanned by the temporal and azimuthal Killing vectors and the one one spanned by the two repeated principal null directions of the Kerr (Petrov type D)  spacetime, aligned with $u_{\rm (car)}\pm e_{\hat r}$.
One may form adapted frames to any test particle world line, e.g. moving along a timelike geodesic,  by conveniently boosting adapted frames to each of them.

\subsection{The static observes and their relative adapted frame}
The static observers, 
which exist only in the spacetime region outside the black hole ergosphere where $g_{tt}<0$,  form a congruence of accelerated, nonexpanding and locally rotating world lines. 
They are, however, nonrotating with respect to observers at rest at spatial infinity and have 4-velocity $u=m$ where
\beq
m=\frac{1}{\sqrt{-g_{tt}}}\,\partial_t
=\left(1-\frac{2Mr}{\Sigma}\right)^{-1/2}\,\partial_t\,.
\eeq
An orthonormal frame adapted to $m$ is
\begin{eqnarray}
\label{static_triad}
e(m)_1 &=&\frac{1}{\sqrt{g_{rr}}}\,\partial_r =\sqrt{\frac{\Delta}{\Sigma}} \,\partial_r
\equiv e_{\hat r}
\,,\nonumber\\ 
e(m)_2 &=&\frac{1}{\sqrt{g_{\theta\theta}}}\,\partial_\theta = \frac{1}{\sqrt{\Sigma}}\,\partial_\theta
\equiv e_{\hat \theta}
\,,\nonumber\\
e(m)_3 &=&\frac{1}{\sqrt{g_{\phi\phi}-{g_{t\phi}{}^2}/{g_{tt}}}}
    \left(\partial_\phi-\frac{g_{t\phi}}{g_{tt}}\partial_t\right) 
\nonumber\\&=& 
\frac{\sqrt{\Delta -a^2 \sin^2\theta}}{\sin \theta \sqrt{\Delta \Sigma}} \left(\partial_\phi-\frac{2Mar\sin^2\theta}{\Delta -a^2 \sin^2\theta} \partial_t  \right)
\,. \nonumber\\
\end{eqnarray}  

\subsection{The ZAMOs and their relative adapted frame}

The ZAMOs are locally nonrotating (but locally rotating in the azimuthal direction in the same sense as the rotation of the black hole) and exist everywhere outside of the outer horizon,
They have 4-velocity $u=n$ where
\begin{eqnarray}\label{ncoord}
n&=&\sqrt{-g^{tt}}\,\left(\partial_t+\frac{g^{t\phi}}{g^{tt}}\partial_\phi\right)
\nonumber\\
&=& \sqrt{\frac{A}{\Delta\Sigma}}\,\left(\partial_t+\frac{2aMr}{A}\partial_\phi\right)
\,,
\end{eqnarray}
where 
\beq
A=(r^2+a^2)^2-a^2\Delta\sin^2\theta\,.
\eeq
The normalized spatial coordinate frame vectors 
\begin{eqnarray}\label{zamotriad}
e(n)_1&=&e_{\hat r}\,, \qquad
e(n)_2=e_{\hat \theta}\,, \nonumber\\
e(n)_3&=&\frac{1}{\sqrt{g_{\phi\phi}}}\,\partial_\phi 
=  \frac{\sqrt{\Sigma}}{\sin\theta \sqrt{A}} \,\partial_\phi
\equiv e_{\hat \phi}
\end{eqnarray}
together with $n$ form an orthonormal adapted frame.
A boost along $e_{\hat \phi}$ maps $n$ into $m$, i.e., 
\beq
m=\gamma(m,n)[n+\nu(m,n)]\,,
\eeq
with relative velocity in the opposite azimuthal direction as the rotation of the black hole associated with the sign of $a$ (resisting the ``dragging of inertial frames")
\beq
\nu(m,n)=-\frac{2Mr}{\Sigma} \,\frac{a \sin\theta }{\sqrt{\Delta}}\,e_{\hat\phi}\,,
\eeq
and associated Lorentz factor $\gamma(m,n)$,
so that ZAMOs and static observers share the same $r$-$\theta$ 2-plane of their local rest spaces.

\subsection{The Carter observers and their relative adapted frame}

The Carter family of observers $u=u_{\rm (car)}$ are geometrically special because their 4-velocity is aligned with the intersection of two geometrically special 2-planes: the one which is the span of the two Killing vectors $\partial_t$ and $\partial_\phi$ and the other spanned by the two principal null directions of the spacetime. This coincidence connects them to the separability of the geodesic equations as well to the alignment of all relevant vectors and tensors in the Kerr spacetime. In particular their relation to the Killing-Yano 2-form allows the solution of the equations of parallel transport found by Marck through successive boosts which isolate the effective spin precession from the various possible boosts of the spherical frame linked to spatial infinity.

The Carter observers are boosted in the opposite azimuthal direction from the static observers compared to the ZAMOs in order to ``comove" with the black hole, their angular velocity at the outer horizon being defined as that of the black hole itself. Their
$4$-velocity $u_{\rm (car)}$ is given by
\begin{eqnarray}
\label{car_obs}
u_{\rm (car)} &=&\frac{r^2+a^2}{\sqrt{\Delta \Sigma}}\left(\partial_t +\frac{a}{r^2+a^2}\,\partial_\phi  \right)\,,\nonumber\\
u_{\rm (car)}^\flat &=& -\sqrt{\frac{\Delta}{\Sigma}}(dt -a \sin^2 \theta \,d\phi)\,,
\end{eqnarray}
the $\flat$ symbol denoting the fully covariant form of any tensor.
Decomposing it with respect to the static observers 
\beq\label{carsta}
u_{\rm (car)}= \gamma(u_{\rm (car)},m) [m+\nu(u_{\rm (car)},m)]\,,
\eeq
leads to the relative velocity 
\beq
\label{nucardef}
\nu(u_{\rm (car)},m) = \frac{a\sin \theta}{\sqrt{\Delta}} e(m)_3\,.
\eeq
Comparing \eqref{ncoord} and \eqref{car_obs} shows that $u_{\rm(car)}$ lies between $n$ and $m$ as claimed above.

A spherical orthonormal frame adapted to $u_{\rm (car)}$ is obtained by using the triad boosted from the either the ZAMO or static observer spherical frame along the azimuthal direction, with 
\begin{eqnarray}
e_1(u_{\rm (car)}) &=& e_{\hat r}\,,\quad 
e_2(u_{\rm (car)}) =e_{\hat \theta}\,, 
\end{eqnarray}
and
\begin{eqnarray}
e_3(u_{\rm (car)}) &=& \frac{a\sin \theta}{\sqrt{\Sigma}}\left(\partial_t +\frac{1}{a\sin^2\theta}\,\partial_\phi  \right)\,,\nonumber\\
e_3(u_{\rm (car)})^\flat  &=& -\frac{a\sin\theta}{\sqrt{\Sigma}}\left(dt -\frac{r^2+a^2}{a} \,d\phi  \right)\,.
\end{eqnarray}
In terms of boost map we have that
\beq
 e(u_{\rm (car)})_\alpha 
= e(m)_\beta B(u_{\rm (car)},m)^\beta{}_\alpha\,.
\eeq

\section{Geodesic observers}

A geodesic timelike world line has a 4-velocity unit tangent vector $U=U^\alpha \partial_\alpha$ with coordinate components $U^\alpha=dx^\alpha/d\tau$ which  can be expressed using the Killing symmetries \cite{Carter:1968ks,Chandrasekhar:1985kt} 
 as a system of first order differential equations
\begin{eqnarray}
\label{geo_eqs}
\frac{d t}{d \tau}&=& \frac{1}{\Sigma}\left[aB+\frac{(r^2+a^2)}{\Delta}P\right]\,,\nonumber \\
\frac{d r}{d \tau}&=&\epsilon_r \frac{1}{\Sigma}\sqrt{R}\,,\nonumber \\
\frac{d \theta}{d \tau}&=&\epsilon_\theta \frac{1}{\Sigma}\sqrt{\Theta}\,,\nonumber \\
\frac{d \phi}{d \tau}&=& \frac{1}{\Sigma}\left[\frac{B}{\sin^2\theta}+\frac{a}{\Delta}P\right]\,,
\end{eqnarray}
where $\tau$ is a proper time parameter along the geodesic,
$\epsilon_r$ and $\epsilon_\theta$ are sign indicators, and
\begin{eqnarray}
\label{geodefs}
P&=& E(r^2+a^2)-L\,a=Er^2-a x\,,\nonumber\\
B&=& L-aE \sin^2\theta=x+aE\cos^2\theta\,, \nonumber\\
R&=& P^2-\Delta (r^2+K)\,,\nonumber\\
\Theta&=&K-a^2\cos^2\theta-\frac{B^2}{\sin^2\theta}\,.
\end{eqnarray}
Here $E$ and $L$ denote the conserved Killing energy and angular momentum per unit mass and $K$ is a separation constant, usually called the Carter constant, while 
the combination $x=L-aE$ proves to be useful. For example, 
in place of $K$ one often uses
\beq
\label{Q_def}
Q= K - (L-aE)^2= K-x^2\,,
\eeq
which vanishes for equatorial plane orbits.
Corresponding to the 4-velocity vector field $U$
is the index-lowered 1-form
\begin{eqnarray}
\label{Ugeogen}
U^\flat &=& -E\, dt +\frac{\Sigma}{\Delta}  \dot r \,dr +\Sigma \dot \theta \,d\theta +L \,d\phi\nonumber\\
&=&
 -E\, dt +\frac{\epsilon_r \sqrt{R(r)}}{\Delta}   \,dr +\epsilon_\theta \sqrt{\Theta(\theta)}\,d\theta +L \,d\phi
\,.\nonumber\\
\end{eqnarray}
Here we use the overdot notation $\dot f=df/d\tau$ for the proper time derivative along the geodesic.
A remarkable (but not very familiar) property of the geodesic family of world lines is that they form an irrotational congruence, $dU^\flat =0$, since each of the covariant component of $U$ depends on the coordinates in a separated form. Consequently,  there exists a new temporal parameter, say  $T$, 
such that
\beq
U^\flat = -dT\,.
\eeq
The relation of $T$ with the Boyer-Lindquist coordinates follows immediately
\beq
T= E t -L\phi -\epsilon_r \int^r\frac{\sqrt{R(r)}}{\Delta}   \,dr -\epsilon_\theta \int^\theta \sqrt{\Theta(\theta)}\,d\theta \,,
\eeq
and can be further expressed in terms of elliptic functions \cite{Bini:2014uua}.

\subsection{Decomposing $U$ in Carter's frame: a new family of radially moving observers $u_{\rm(rad)}$}

Consider a generic timelike geodesic with unit tangent vector \eqref{Ugeogen} decomposed relative to the Carter observers.
To make the notation less cumbersome below we introduce the abbreviations $\gamma(U,u_{\rm (car)})=\gamma_{\rm c}$, $\nu(U,u_{\rm (car)})=\nu_{\rm c}$.
For later use let us introduce the angular part $\nu^\top$ of the Carter relative velocity and an orthogonal vector $\nu^\perp=  e(u_{\rm (car)})_1 \times_{u_{\rm (car)}} \nu^\top $ of the same magnitude in the angular subspace
\begin{eqnarray}
\nu^\top &=& \nu_{\rm c}^2  e(u_{\rm (car)})_2+ \nu_{\rm c}^3  e(u_{\rm (car)})_3\equiv ||\nu^\top ||  \hat \nu^\top\,, \nonumber\\
\nu^\perp &=& -\nu_{\rm c}^3  e(u_{\rm (car)})_2+\nu_{\rm c}^2  e(u_{\rm (car)})_3\equiv ||\nu^\perp ||  \hat \nu^\perp\,, \nonumber\\
\end{eqnarray}
with $||\nu^\top ||=||\nu^\perp || =\sqrt{(\nu_{\rm c}^2)^2+(\nu_{\rm c}^3)^2}$.
We will use the notation
\begin{eqnarray}
\label{nu_top_perp}
\hat \nu^\top &=& \cos \Phi\, e(u_{\rm (car)})_2-\sin \Phi\,  e(u_{\rm (car)})_3\,,\nonumber\\
\hat \nu^\perp &=& \sin \Phi\, e(u_{\rm (car)})_2 +\cos \Phi\,  e(u_{\rm (car)})_3\,,
\end{eqnarray}
with
\beq
\label{Phi_rot}
\tan \Phi = - \frac{\nu_{\rm c}^3 }{\nu_{\rm c}^2 }\,.
\eeq
The orthogonal decomposition into radial and angular directions in the Carter frame is the starting point for solving the equations of parallel transport along a geodesic, as done by Marck and explained geometrically for the simpler case of equatorial plane geodesics in our previous articles
\cite{Bini:2016iym,Bini:2016ovy,Bini:2017slb}. 

Using the components of $U$ with respect to the Carter observers we define a frame $\{e(u_{\rm(rad)})_\alpha\}$ (with $e(u_{\rm(rad)})_0\equiv u_{\rm(rad)}$) in which we first boost along the radial direction by the radial component of the relative 4-velocity of the gyro to obtain a new radially comoving radial direction in a new local rest space, and then pick the next frame vector to be along the direction of the remaining angular component of the Carter relative velocity,  and then the final angular axis orthogonal to the first one forming a right-handed spatial frame within $LRS_{u_{\rm(rad)}}$
\begin{eqnarray}
\label{rad_frame}
u_{\rm(rad)} &=&\gamma^\Vert [u_{\rm (car)}+\nu_{\rm c}^1 e(u_{\rm (car)})_1 ]\,,\nonumber\\
&=& \cosh \alpha u_{\rm (car)}+\sinh \alpha  e(u_{\rm (car)})_1\nonumber\\
e(u_{\rm(rad)})_1 &=&\gamma^\Vert [\nu_{\rm c}^1 u_{\rm (car)}+ e(u_{\rm (car)})_1]\nonumber\\
&=& \sinh \alpha u_{\rm (car)}+\cosh \alpha  e(u_{\rm (car)})_1\nonumber\\
e(u_{\rm(rad)})_2 &=& \hat \nu^\top  \nonumber\\
&=& \cos \Phi \,e(u_{\rm (car)})_2-\sin \Phi \,e(u_{\rm (car)})_3 ,\nonumber\\
e(u_{\rm(rad)})_3 &=& \hat \nu^\perp\nonumber\\
&=& \sin \Phi \,e(u_{\rm (car)})_2+\cos \Phi \,e(u_{\rm (car)})_3\,,
\end{eqnarray}
where the rotation angle $\Phi$ is defined in Eq.~\eqref{Phi_rot} and the boost rapidity $\alpha$ is defined so that
\beq
\label{nu_c1}
\nu_{\rm c}^1=\tanh\alpha\,,\qquad
\gamma^\Vert=\cosh\alpha
=\frac{P}{\sqrt{\Delta(r^2+K)}}\,.
\eeq
In compact form we have
\beq
 e(u_{\rm (rad)})_\alpha 
= e(u_{\rm (car)})_\beta B(u_{\rm (rad)},u_{\rm (car)})^\beta{}_\alpha\,,
\eeq
with $B(u_{\rm (rad)},u_{\rm (car)})=\tilde B(u_{\rm (rad)},u_{\rm (car)}) R_{23}(\Phi)=R_{23}(\Phi)\tilde B(u_{\rm (rad)},u_{\rm (car)}) $, where
\beq
\label{B_urad_u_car_b}
(\tilde B(u_{\rm (rad)},u_{\rm (car)})^\gamma{}_\epsilon )=
\begin{pmatrix}
\cosh\alpha&\sinh\alpha&0&0\cr
\sinh\alpha&\cosh\alpha&0&0\cr
0&0&1&0\cr
0&0&0&1\cr
\end{pmatrix}
\,,
\eeq
and
\beq
\label{B_urad_u_car_r}
(R_{23}(\Phi)^\alpha{}_\beta )=
\begin{pmatrix}
1&0&0&0\cr
0&1&0&0\cr
0&0&\cos\Phi&\sin\Phi\cr
0&0&-\sin\Phi&\cos\Phi\cr
\end{pmatrix}
\,.
\eeq
We see that the radially moving observers have an adapted frame which is the boost of the Carter frame onto their own LRS plus a spatial rotation in the 2-plane $e(u_{\rm (car)})_2-e(u_{\rm (car)})_3$. 
One could have defined a rotated Carter frame 
\begin{eqnarray}
e'(u_{\rm (car)})_0&=&u_{\rm (car)}\,,\quad
e'(u_{\rm (car)})_1=e(u_{\rm (car)})_1\,,\nonumber\\
e'(u_{\rm (car)})_2&=&\cos\Phi\,e(u_{\rm (car)})_2-\sin\Phi\,e(u_{\rm (car)})_3\,,\nonumber\\
e'(u_{\rm (car)})_3&=&\sin\Phi\,e(u_{\rm (car)})_2+\cos\Phi\,e(u_{\rm (car)})_3\,,
\end{eqnarray}
or in compact form
\beq
\label{eucarprime}
e'(u_{\rm (car)})_\sigma=e(u_{\rm (car)})_\gamma R_{23}(\Phi)^\gamma{}_\sigma\,.
\eeq
In that case the radially moving observer adapted frame is simply a boost of the frame $\{e'(u_{\rm (car)})_\alpha\}$.

The (timelike) geodesic 4-velocity $U$ in this frame then has the   form  
\beq\label{Ubetarad}
U=\cosh \beta\, u_{\rm(rad)}+\sinh \beta\, e(u_{\rm(rad)})_2\,,
\eeq
where
\beq
\label{beta_def}
\cosh \beta =\sqrt{\frac{K+r^2 }{\Sigma}}\,,\qquad \sinh \beta =  \sqrt{\frac{K-a^2\cos^2\theta }{\Sigma}}\,.
\eeq
From this relation one easily identify the orthogonal (spatial) direction in this plane 
\beq\label{e1def}
e(U)_3=\sinh \beta\, u_{\rm(rad)}+\cosh \beta\, e(u_{\rm(rad)})_2\,.
\eeq

Marck showed that a unit vector $e(U)_2$ orthogonal to both $U$ and $e(U)_3$ which is also parallel propagated along $U$ arises naturally by normalizing the electric part $f_{\alpha\beta} U^\beta$ of the Killing-Yano 2-form $f$ of the Kerr spacetime with respect to $U$. Because this 2-form is so simply expressed in both the Carter and intermediate frames (see Appendix A of Ref.~\cite{Bini:2016iym}), the resulting vector frame components
are obtained by a simple anisotropic rescaling of the two vector components of $U$ expressed in the form \eqref{Ubetarad}
\begin{eqnarray}
\label{rot12}
e(U)_2
&=&-\sin\Xi\, e(u_{\rm(rad)})_1+\cos\Xi\, e(u_{\rm(rad)})_3
\,,
\end{eqnarray}
where
\beq
\label{Xi_def}
  \cos\Xi= \frac{r}{\sqrt{K}}\,\sinh\beta\,,\quad
  \sin\Xi= \frac{a\cos \theta}{\sqrt{K}}\,\cosh\beta\,.
\eeq
The last frame vector $e(U)_1=e(U)_2\times_U e(U)_3$
 is then determined by orthogonality to be
\begin{eqnarray}
e(U)_1
&=& \cos\Xi\, e(u_{\rm(rad)})_1+\sin\Xi\, e(u_{\rm(rad)})_3\,.
\end{eqnarray}

\section{Boosted frames}

The construction of the various natural adapted frames along timelike geodesics is facilitated by using the boost maps introduced above (see, e.g., Eq.~\eqref{Boost_expl}) applied to the various spherical frames associated with the Boyer-Lindquist coordinates, that is
\beq
e(U,u)_a = B(U,u)e(u)_a\,,
\eeq
for any observer family $u=m,n,u_{\rm (car)},u_{\rm (rad)}$, namely
\begin{eqnarray}
e(U,m)_a &=& B(U,m)e(m)_a\,,\nonumber\\
e(U,n)_a &=& B(U,n)e(n)_a\,,\nonumber\\
e(U,u_{\rm (car)})_a &=& B(U,u_{\rm (car)})e(u_{\rm (car)})_a\,,\nonumber\\
e(U,u_{\rm (rad)})_a &=& B(U,u_{\rm (rad)}) e(u_{\rm (rad)})_a\,.
\end{eqnarray}
They are related to each other by Wigner rotations in $LRS_U$.
One can study their parallel transport along $U$, defining corresponding angular velocities by
\beq
\label{OmegainU}
\frac{D}{d\tau_U} e(U,m)_a= \Omega(U,m) \times_U e(U,m)_a
\eeq 
and similar relations for $\Omega(U,n) $, $\Omega(U,u_{\rm (car)})$ and $\Omega(U,u_{\rm (rad)})$.

One may evaluate the angular velocity $\Omega(U,u)=\Omega(U,u)^a\, e(U,u)_a$ in terms of the relative motion of the geodesic $U$ and the observer family $u$ ($u=m,n,u_{\rm (car)},u_{\rm (rad)}$) as a sum of the following three terms as given in  Ref.~\cite{Jantzen:1992rg}
\beq
\label{OmegaUu}
\Omega(U,u)=-\gamma(U,u) B(U,u)
[\omega_{({\rm fw},u)}+\omega_{({\rm sc},U,u)}+\omega_{({\rm geo},U,u)}]\,,
\eeq
using its notation for the Fermi-Walker and the spatial curvature angular rotation vectors which characterize the covariant derivatives of the orthonormal frame along the orbit
\beq \label{omegainu}
P(u)\nabla_U e(u)_a = -\gamma(U,u)[\omega_{({\rm fw},u)}
+\omega_{({\rm sc},U,u)}]\times_{u}e(u)_a\,,\nonumber
\eeq
as well as the geodetic precession term in the gyroscope precession formula (see Eq.~(9.10) of 
Ref.~\cite{Jantzen:1992rg})
\beq
\label{omgUu}
\omega_{({\rm geo},U,u)}=\frac{1}{1+\gamma(U,u) } \,\nu(U,u)\times_u F^{(G)}_{({\rm fw},U,u)}\,,
\eeq
defined in terms of the spatial gravitational force 
$ F^{(G)}_{({\rm fw},U,u)} = -\nabla_U \,u$.
Additional details, including notation, can be found in Ref.~\cite{Jantzen:1992rg} and will not be repeated here.

\section{Spin precession frames}

In a stationary spacetime one can formulate a precise way of measuring locally the precession of the spin of a test gyroscope in geodesic motion with respect to the ``distant stars," namely the celestial sphere at spatial infinity \cite{Bini:2016iym,Bini:2016ovy,Bini:2017slb}. The static observers determine a local reference frame rigidly linked to the distant celestial sphere in the sense that these observers see an unchanging distant sky pattern of incoming photons from spatial infinity. Thus one can establish a Cartesian frame in the local rest space along each static observer world line which establishes the local celestial sky, and the Killing symmetry links that frame to a single Cartesian frame at spatial infinity. With the additional axial symmetry and reflection symmetry across the equatorial plane, the natural Cartesian frame linked to the Boyer-Lindquist spherical frame is uniquely a candidate for this link, but off the equatorial plane of the Kerr spacetime, though not unique (in the case of non-spherical symmetry), it is the simplest frame to serve this purpose. Spheroidal spatial coordinates would provide another spherical orthonormal triad which could be used to introduce a corresponding Cartesian frame, for example, that would differ from the Boyer-Lindquist frame except on the equatorial plane.

For a gyro at rest with respect to the static observer grid, the precession is unambiguous and straightforward to measure, but for relative motion one has the additional complication that in the local rest space of the gyro, the local axes linked to spatial infinity by the incoming null geodesics from spatial infinity in the static observer local rest space are distorted by spatial aberration. Boosting those axes into the local rest space of the gyro enables one to measure the relative rotation of the spin vector, or equivalently, boosting the spin vector back to the static observer rest frame gives the same result (because of the isometric property of the boost). Of course one can compare the projection of the spin vector into the static observer rest space, the spin vector as seen by that observer, but even for circular motion in flat spacetime, this projected spin vector does not undergo a uniform precession, instead undergoing a periodic tilting (in spacetime) effect to maintain orthogonality with the 4-velocity of the gyro resulting in a changing spin vector magnitude and orbital-dependent additional rotation. The traditional Thomas precession formula for circular orbits, for example, describes this boosted spin vector precession \cite{TW,circles}.

The Carter spatial frame is azimuthally boosted with respect to the static observer spherical frame, which in turn is azimuthally boosted with respect to the Boyer-Lindquist spherical coordinate normalized frame. The Carter frame is associated with the Killing tensor and Killing 2-form that in turn are associated with the separability of the geodesic equations, and the solution of the equations for Fermi-Walker transport along those geodesics, the latter of which describes free fall gyro spin behavior in the timelike case. The Marck procedure for solving the differential equations for a parallely transported frame along geodesics (which is also a Fermi-Walker frame in this case) relies on a two step procedure for transforming to the local rest frame of the geodesic, first boosting the Carter frame along the radial direction common to all three frames (ZAMO, static, Carter) to comove radially with the gyro, preserving the Killing 2-form in this step, then followed by a boost in the remaining angular direction to comove with the gyro. This defines his preliminary frame $e_{\rm(mar)}(U)_\alpha$ which isolates the remaining parallel transport rotation to a 2-plane of the first and last spatial frame vectors, leaving the remaining frame vector aligned with its spatial normal direction invariant under parallel transport.
In fact, we have 
\begin{eqnarray}
e_{\rm(mar)}(U)_\alpha&=&e(U,u_{\rm (rad)})_\epsilon R_{12}(\Xi)^\epsilon{}_\alpha\nonumber\\
&=& e(u_{\rm (rad)})_\epsilon R_{12}(\Xi)^\epsilon{}_\beta
 B(U,u_{\rm (rad)})^\beta{}_\alpha\,,\nonumber\\
\end{eqnarray}
where\footnote{
The vector $e(U)_2$ defined in Eq.~\eqref{rot12} has an overall sign in front with respect to the definition used in Ref.~\cite{Bini:2016iym}. This implies that the matrix $R_{12}(\Xi)$ is simply a rotation instead of being the composition of a rotation and a reflection.
}

\begin{widetext}

\beq
(R_{12}(\Xi)^\epsilon{}_\beta )=
\begin{pmatrix}
1&0&0&0\cr
0&\cos\Xi&\sin\Xi&0\cr
0&-\sin\Xi&\cos\Xi&0\cr
0&0&0&1\cr
\end{pmatrix}
\,,\qquad
(B(U,u_{\rm (rad)})^\beta{}_\alpha )=
\begin{pmatrix}
\cosh\beta&0&0&\sinh\beta\cr
0&1&0&0\cr
0&0&1&0\cr
\sinh\beta&0&0&\cosh\beta\cr
\end{pmatrix}
\,.
\eeq
Therefore,
\begin{eqnarray}
 e_{\rm(mar)}(U)_\alpha
&=&  e(u_{\rm (rad)})_\epsilon R_{12}(\Xi)^\epsilon{}_\beta
 B(U,u_{\rm (rad)})^\beta{}_\alpha
\nonumber\\
&=&   e(u_{\rm (car)})_\gamma B(u_{\rm (rad)},u_{\rm (car)})^\gamma{}_\epsilon 
R_{12}(\Xi)^\epsilon{}_\beta
B(U,u_{\rm (rad)})^\beta{}_\alpha
\nonumber\\
&=&   e(u_{\rm (car)})_\gamma B(u_{\rm (rad)},u_{\rm (car)})^\gamma{}_\epsilon
 B(U,u_{\rm (rad)})^\epsilon{}_\beta
R_{12}(\Xi)^\beta{}_\alpha
\,,
\end{eqnarray}
with $B(u_{\rm (rad)},u_{\rm (car)})=\tilde B(u_{\rm (rad)},u_{\rm (car)}) R_{23}(\Phi)$ given in Eqs.~\eqref{B_urad_u_car_b} and \eqref{B_urad_u_car_r}\,.
We have then
\begin{eqnarray}
 e_{\rm(mar)}(U)_\alpha
&=&   e(u_{\rm (car)})_\gamma R_{23}(\Phi)^\gamma{}_\sigma \tilde B(u_{\rm (rad)},u_{\rm (car)})^\sigma{}_\epsilon
 B(U,u_{\rm (rad)})^\epsilon{}_\beta
R_{12}(\Xi)^\beta{}_\alpha
\,.
\end{eqnarray}
Using then the results of Section III we find
\beq
\tilde B(u_{\rm (rad)},u_{\rm (car)})^\sigma{}_\epsilon
 B(U,u_{\rm (rad)})^\epsilon{}_\beta=\hat B(U,u_{\rm (rad)},u_{\rm (car)})^\sigma{}_\epsilon
R^{\rm (W)}(U,u_{\rm (rad)},u_{\rm (car)})^\epsilon{}_\beta\,,
\eeq
with
\beq
\hat B(U,u_{\rm (rad)},u_{\rm (car)})^\sigma{}_\epsilon= 
\left.{\mathcal B}({\pmb \nu_3})^\alpha{}_\gamma\right|_{\alpha_1=\alpha,\alpha_2=\beta}
\eeq
given by \eqref{Bnu3},
and
\beq
(R^{\rm (W)}(U,u_{\rm (rad)},u_{\rm (car)})^\epsilon{}_\beta)
= 
\begin{pmatrix}
1 &0 &0 &0  \cr
0 &\cos\theta & 0& \sin\theta \cr
0 &0 &1 &0  \cr
0 &-\sin\theta &0& \cos\theta  \cr
\end{pmatrix}
\,,
\eeq
with 
\beq
\cos\theta=\frac{\cosh\alpha+\cosh\beta}{1+\cosh\alpha\cosh\beta}\,,\qquad
\sin\theta=\frac{\sinh\alpha\sinh\beta}{1+\cosh\alpha\cosh\beta}\,.
\eeq
This implies 
\begin{eqnarray}
 e_{\rm(mar)}(U)_\alpha
&=&   e(u_{\rm (car)})_\gamma R_{23}(\Phi)^\gamma{}_\sigma \hat B(U,u_{\rm (rad)},u_{\rm (car)})^\sigma{}_\epsilon
R^{\rm (W)}(U,u_{\rm (rad)},u_{\rm (car)})^\epsilon{}_\beta
R_{12}(\Xi)^\beta{}_\alpha
\,,
\end{eqnarray}
that is by using Eq.~\eqref{eucarprime}
\begin{eqnarray}
 e_{\rm(mar)}(U)_\alpha
&=&   e'(u_{\rm (car)})_\sigma \hat B(U,u_{\rm (rad)},u_{\rm (car)})^\sigma{}_\epsilon
R^{\rm (W)}(U,u_{\rm (rad)},u_{\rm (car)})^\epsilon{}_\beta
R_{12}(\Xi)^\beta{}_\alpha
\,.
\end{eqnarray}

\end{widetext}

One can pass then from $u_{\rm (car)}$ to $m$ by using a pure boost
\beq
(B(u_{\rm (car)},m)^\gamma{}_\epsilon )=
\begin{pmatrix}
\cosh\delta&0&0&\sinh\delta\cr
0&1&0&0\cr
0&0&1&0\cr
\sinh\delta&0&0&\cosh\delta\cr
\end{pmatrix}
\,,
\eeq
where $\gamma(u_{\rm (car)},m)=\cosh\delta$ and $\nu(u_{\rm (car)},m)=\tanh\delta=a\sin\theta/\sqrt{\Delta}$ (see Eq.~\eqref{nucardef}). 

This frame $\{U,e_{(\rm mar)}(U)_a\}$ is a degenerate Frenet-Serret frame along $U$, such that
\begin{eqnarray}\label{FSframeDEs}
&&\frac{DU}{d\tau}= 0\,,\
\frac{De_{(\rm mar)}(U)_1}{d\tau} = {\mathcal T} e_{(\rm mar)}(U)_3\,,\\
&&\frac{De_{(\rm mar)}(U)_2}{d\tau} = 0 \,,\
\frac{De_{(\rm mar)}(U)_3}{d\tau} = -{\mathcal T} e_{(\rm mar)}(U)_1\,,\nonumber
\end{eqnarray}
with 
\begin{eqnarray}
\label{cal_T_torsion}
{\mathcal T}&=&\frac{\sqrt{K}}{\Sigma}\left[ \frac{P}{r^2+K}+\frac{aB}{K-a^2\cos^2\theta} \right]\nonumber\\
&=& \frac{\sqrt{K}}{(r^2+K)(K-a^2\cos^2\theta)}(KE+ax)\nonumber\\
&=& -\sqrt{K\Sigma} \left[\frac{\sqrt{\Delta}}{r^2+K} u_{\rm (car)}\right.\nonumber\\
&&\left. -\frac{a\sin \theta}{K-a^2\cos^2\theta}e(u_{(\rm car)})_3 \right]\cdot U\,,\nonumber
\end{eqnarray}
the only surviving (spacetime) torsion of the world line, which reversed in sign describes the scalar angular velocity of two frame vectors with respect to parallel transport. [Here $r$ and $\theta$ actually mean $r(\tau)$ and $\theta(\tau)$] The vector angular velocity of the frame is aligned with the parallel transported third spatial direction in this frame.
\beq\label{omegaprec}
  \Omega_{\rm par} = -\mathcal T   e_{(\rm mar)}(U)_2
\,.
\eeq
Thus a parallel transported frame is obtained by further rotating this pair of frame vectors along the world line at the opposite angular velocity $d\Psi/d\tau = \mathcal T $
\beq
   \left(\begin{array}{c} e_{(\rm par)}(U)_{1}\\e_{(\rm par)}(U)_{3} \end{array} \right)
= \left(\begin{array}{cc} \cos \Psi &  -\sin \Psi \\ \sin \Psi &  \cos \Psi  \end{array} \right)
   \left(\begin{array}{c}  e_{(\rm mar)}(U)_{1}\\  e_{(\rm mar)}(U)_{3} \end{array} \right)
\,,
\eeq
while the third vector remains unchanged
\beq
e_{(\rm par)}(U)_{2}= e_{(\rm mar)}(U)_{2} \,.
\eeq

The spin vector of a test gyro moving along such a geodesic simply has constant components in this final parallel transported frame. However, to compare this evolution with the (nonrotating) celestial sphere at spatial infinity, one needs the Marck frame which is anchored to the local spherical frame modulo boosts.
The Marck frame vector $e_{\rm(mar)}(U)_{1}$ is locked to the radial direction $e_{\hat r} $  in the spherical grid of the static observers following the time lines, 
differing only by a boost due to the radial motion of the gyro alone.
Along the gyro geodesic world line, the spherical axes rotate with an orbital angular velocity with respect to spatial infinity, so one must subtract this more complicated rotation from the simpler parallel transport rotation to obtain the net rotation of the axes with respect to spatial infinity. For equatorial plane motion this is simple since both the orbital and parallel transport rotations lie in a 2-plane and it is a matter of subtracting the two scalar angular velocities to get the net angular velocity of precession in the Marck frame. These matters are discussed in \cite{Bini:2016iym,Bini:2016ovy,Bini:2017slb}.

\subsection{The equatorial plane limit}

For equatorial geodesics the above discussion simplify notably. Indeed, the condition $\theta=\pi/2$ implies  $U^\theta=0$ and, from Eq.~\eqref{Q_def},
\beq
K=x^2\,.
\eeq
In turn, from Eq.~\eqref{nu_c1}, we have
\beq 
\cosh\alpha = \frac{P}{\sqrt{\Delta (r^2+x^2)}}\,.
\eeq
Moreover $\nu_{\rm c}^2=0$ so that,  from Eq.~\eqref{nu_top_perp}, it follows that
$ \Phi={\pi}/{2}$
and the rotation $R_{23}(\Phi)$ reduces to
\beq
\label{B_urad_u_car_r_equat}
(R_{23}\left({\pi}/{2}\right)^\alpha{}_\beta )
=
\begin{pmatrix}
1&0&0&0\cr
0&1&0&0\cr
0&0&0&1\cr
0&0&-1&0\cr
\end{pmatrix}
\,.
\eeq
Finally, from Eq.~\eqref{beta_def}, we find
\beq
\sinh \beta =\frac{|x|}{r}
\eeq
and from Eq.~\eqref{Xi_def}
\beq
\Xi =0\,.
\eeq
Summarizing, the geodesic 4-velocity writes as in Eq.~\eqref{Ubetarad} and an adapted (spatial) frame to $U$ has
$e(U)_3$ given by Eq.~\eqref{e1def} and
\beq
e(U)_1=e(u_{\rm rad})_1\,,\qquad
e(U)_2=-e(u_{\rm rad})_3\,.
\eeq
In matrix form
\beq
e(U)_\alpha =e(u_{\rm rad})_\beta {\mathcal A}^\beta{}_\alpha\,,
\eeq
with
\beq
({\mathcal A}^\beta{}_\alpha )=
\begin{pmatrix}
\cosh\beta & 0 &\sinh\beta &0\cr
0&1&0& 0\cr
0&0&0&1\cr
\sinh\beta &0 & \cosh\beta &0\cr
\end{pmatrix}
\begin{pmatrix}
1 & 0 &0 &0\cr
0&1&0& 0\cr
0&0&1&0\cr
0&0 & 0 &-1\cr
\end{pmatrix}
\,,
\eeq
which is the product of a boost $B(U,u_{\rm rad})$ along $e(u_{\rm rad})_2$ and a \lq\lq reflection" across  $e(u_{\rm rad})_3$.

In turn, from Eq.~\eqref{rad_frame} the \lq\lq rad" frame is simply related to Carter's frame as
\begin{eqnarray}
\label{rad_frame2}
u_{\rm(rad)} 
&=& \cosh \alpha u_{\rm (car)}+\sinh \alpha  e(u_{\rm (car)})_1\,,\nonumber\\
e(u_{\rm(rad)})_1 
&=& \sinh \alpha u_{\rm (car)}+\cosh \alpha  e(u_{\rm (car)})_1\,,\nonumber\\
e(u_{\rm(rad)})_2
&=& -e(u_{\rm (car)})_3 ,\nonumber\\
e(u_{\rm(rad)})_3 &=&  \,e(u_{\rm (car)})_2\,,
\end{eqnarray}
or, in matrix form
\beq
e(u_{\rm rad})_\alpha = e(u_{\rm car})_\beta {\mathcal W}^\beta{}_\alpha\,,
\eeq
with
\beq
({\mathcal W}^\beta{}_\alpha )
= \begin{pmatrix}
\cosh\alpha &\sinh\alpha &0&0\cr
\sinh\alpha &\cosh\alpha &0&0\cr
0&0&1&0\cr
0&0&0&1\cr
\end{pmatrix}
\begin{pmatrix}
1& 0 &0&0\cr
0&1&0& 0\cr
0&0&0&-1\cr
0 &0 & 1 &0\cr
\end{pmatrix}
\,.
\eeq
where the first matrix is $\tilde B(u_{\rm (rad)},u_{\rm (car)}) $ given by
Eq.~\eqref{B_urad_u_car_b} and the second is $R_{23}(\pi/2)$.
Combining the two operations leads to 
\beq
e(U)_\alpha = e(u_{\rm car})_\mu {\mathcal W}^\mu{}_\beta {\mathcal A}^\beta{}_\alpha\,.
\eeq
Finally, the torsion ${\mathcal T}$ reduces to
\begin{eqnarray}
\label{cal_T_torsion}
{\mathcal T}&=&\frac{|x|}{r^2+x^2}\left( E+\frac{a}{x}\right)\,.
\end{eqnarray}
Note that the sign of $x=L-aE$ depends on the relative signs of $L$ and $a$, i.e., if the orbit is either prograde (same sign) or retrograde (opposite sign).
For instance, if $a>0$ then $x>0$ ($x<0$) for prograde (retrograde) orbits.

\subsection{Bypassing Marck}

One could resolve the parallel transport equations  without relying on the Killing-Yano 2-form essential for Marck's elegant construction by a brute force approach. The complication is that one cannot confine the essential rotation to a parallel transported 2-plane, but would require a general rotation parametrized by three angles (two to determine the axis of rotation and one for the rotation about this axis)  and a vector angular velocity rather than a single scalar one. One simply boosts the static spherical frame into $LRS_U$ and evaluates the parallel transport angular velocity of the boosted axes. Such a direct calculation was done in Ref. \cite{Bini:2017slb} and is not so complicated, but lacks the geometrical interpretation of Marck's construction. Such an approach could be extended naturally to any spacetime lacking a Killing-Yano 2-form, but for which the geodesic equations are separable, allowing this frame to be expressed in terms of the constants of the motion. 

\section{Isolating cumulative precession effects}

We now discuss the cumulative precession effects on a test gyroscope moving along an equatorial plane geodesic orbit. While the bound case is been given much more attention in the literature and explicit expressions already exist describing the precession \cite{Akcay:2016dku,Akcay:2017azq,Kavanagh:2017wot,Bini:2018aps}, the unbound case has only been studied within certain approximation schemes, namely in a post-Newtonian expansion (weak-field and slow-motion, in a power series in the reciprocal of the speed of light, $1/c$) and  in a post-Minkowskian expansion (weak-field, power series in the gravitational constant $G$) \cite{Damour:2016gwp,Bini:2017wfr,Damour:2017zjx,Bini:2018ywr,Bini:2017ldh,Bini:2017pee,Bini:2018zxp}. 

We evaluate exactly the total spin-precession angle $\Psi$, the accumulated azimuthal phase $\Phi$ and the associated spin precession invariant \cite{Dolan:2013roa}
\beq
\psi=1-\frac{\Psi}{\Phi}\,.
\eeq
For bound orbits these quantities are evaluated between two successive passages at periastron  corresponding to one period of the radial motion, whereas for unbound (hyperbolic-like) orbits they are evaluated for the entire scattering process (from the two asymptotic states at spatial infinity).
We provide in both cases closed form analytical expressions in terms of elliptic functions as well as approximate expressions which facilitate comparison with known results. The combined quantity $\psi$ is a sort of average azimuthal precession rate, since the full spin precession with respect to spatial infinity measured from the perihelion is just $\psi\,\Phi$.

\subsection{Gyroscope moving along a bound equatorial orbit}

In the case of bound equatorial orbits not captured by the black hole, the radial motion is periodic and confined between a minimum radius $r_{\rm per}$ (periastron) and a maximum radius $r_{\rm apo}$ (apastron). 
It is convenient to introduce the relativistic anomaly $\chi\in[0,2\pi]$ such that 
\beq\label{rdichi}
r=\frac{Mp}{1+e\cos\chi}\,,
\eeq
with dimensionless semi-latus rectum $p=1/u_p$ and eccentricity $0\leq e<1$ (see, e.g., Ref. \cite{Bini:2016iym} for additional details).
The parameters $(u_p,e)$ are related, in turn, to the energy and angular momentum (per unit mass) $(E,L)$ entering the geodesic equation \eqref{geo_eqs}, with $E<1$. This relation \eqref{rdichi} with these parameters represents a classical Newtonian orbit in the Boyer-Lindquist polar coordinates in the equatorial plane  which precesses due to general relativity according to a function $\chi$ rather than the azimuthal angle $\phi$ itself.

In terms of $\chi$, the rate of gyroscope precession and azimuthal change are given by

\begin{widetext}

\begin{eqnarray}
\label{dphidchi}
\frac{d\phi}{d\chi}&=& u_p^{1/2}\frac{ \hat x + \hat a E - 2 u_p \hat x (1+  e\cos \chi) }{[1+u_p^2 \,\hat x{}^2 (e^2-2 e\cos \chi -3) ]^{1/2}
[1-2 u_p(1+ e\cos \chi) +\hat a^2 u_p^2(1+ e\cos \chi)^2  ]}
\,,\nonumber\\
\frac{d\Psi}{d\chi}&=&u_p^{1/2}\frac{\hat a+E\hat x}{[1-u_p^2\hat x^2(3-e^2+2e\cos\chi)]^{1/2}[1+u_p^2\hat x^2(1+e\cos\chi)^2]}\,,
\end{eqnarray}
where ${d\Psi}/{d\tau}={\mathcal T}$ is defined in Eq.~\eqref{cal_T_torsion} and
\beq
M \frac{d\chi}{d\tau} =  u_p^{3/2}(1+e\cos \chi )^2
[1+u_p^2\, \hat x{}^2 ( e^2-2 e\cos\chi-3)]^{1/2}\,.
\eeq
Eqs.~\eqref{dphidchi}, once integrated over a radial period, i.e., 
\beq
\Phi=\int^{2\pi}_0\frac{d\phi}{d\chi}d\chi\,,\qquad
\Psi=\int^{2\pi}_0\frac{d\Psi}{d\chi}d\chi\,,
\eeq
then leads to $\Phi$, $\Psi$ and $\psi$ being expressible in terms of elliptic functions. 
We find
\begin{eqnarray}
\Phi&=&-\frac{\kappa}{\hat a^2e^2u_p^2\sqrt{eu_p\hat x^2}(b_+-b_-)}\left\{
[\hat L-2u_p\hat x(1+eb_+)]k_+\Pi\left(k_+,\kappa\right)
-[\hat L-2u_p\hat x(1+eb_-)]k_-\Pi\left(k_-,\kappa\right)
\right\}
\,,\nonumber\\
\Psi&=&-\frac{i}{2}\frac{\kappa(\hat a +E\hat x)}{\hat x^2(eu_p)^{3/2}}\left[k\Pi(k,\kappa)-\bar k\Pi(\bar k,\kappa)\right]\,,
\end{eqnarray}
where $\hat L=L/M$,
\beq
\kappa^2=\frac{4eu_p^2\hat x^2}{1-(1-e)(3+e)u_p^2\hat x^2}
\,, \quad
k_\pm=\frac{2}{1+b_\pm}
\,,\quad
b_\pm=\frac{1-\hat a^2u_p\pm\sqrt{1-\hat a^2}}{\hat a^2eu_p}
\,,\quad
k=2ieu_p\hat x\frac{1+i(1-e)u_p\hat x}{1+(1-e)^2u_p^2\hat x^2}
\,,
\eeq
the overbar denoting complex conjugation, and where  
\beq
\Pi(n,m)=\int_0^{\frac{\pi}{2}}\frac{dz}{(1-n\sin^2z)\sqrt{1-m^2\sin^2z}}
\eeq
is the complete elliptic integral of the third kind~\cite{Gradshteyn}.

Expanding these expressions in terms of the eccentricity $e$, so that  $\Phi  = \Phi_0  + e^2 \Phi_{e^2}   +O(e^4)$ and similarly for $\Psi$ and $\psi$, one finds
\begin{eqnarray}
\frac{\Phi}{2\pi} &=&\frac{1}{\sqrt{1-6u_p+8\hat a u_p^{3/2}-3\hat a^2 u_p^2}} \nonumber\\
&+&e^2 \frac{3u_p^2 (-1+2u_p+(-3+22u_p)\sqrt{u_p}\hat a -33u_p^2\hat a^2 +13u_p^{5/2}\hat a^3)(\hat a \sqrt{u_p}-1)^3}{4(1-2u_p+\hat a^2 u_p^2)(1-6u_p+8\hat a u_p^{3/2}-3\hat a^2 u_p^2)^{5/2}}+O(e^4)\,,\nonumber\\
\frac{\Psi}{2\pi} &=& \frac{\sqrt{1-3u_p+2\hat a u_p^{3/2}}}{\sqrt{1-6u_p+8\hat a u_p^{3/2}-3\hat a^2 u_p^2}}+e^2\frac{3P(u_p,\hat a)(\hat a \sqrt{u_p}-1)^2u_p^2}{4(1-2u_p+\hat a^2 u_p^2)(1-6u_p+8\hat a u_p^{3/2}-3\hat a^2 u_p^2)^{5/2}}+O(e^4)\,,
\end{eqnarray}
where
\begin{eqnarray}
P(u_p,\hat a)&=&14\hat a^5 u_p^{9/2}+(-81 u_p +13 )u_p^3\hat a^4+4 u_p^{5/2}(47 u_p-12)\hat a^3+(-225 u_p^2-3+68 u_p) u_p\hat a^2\nonumber\\
&&+2 u_p^{1/2}(72 u_p^2-22 u_p+1)\hat a-1-42 u_p^2+15 u_p\,.
\end{eqnarray}
Finally
\begin{eqnarray}
\psi&=& 1-\sqrt{1-3 u_p+2 \hat a u_p^{3/2}}\nonumber\\
&&-\frac32 e^2\frac{(2\hat a^3 u_p^{5/2}-3\hat a^2 u_p^2-2\hat a u_p^{3/2}+4 u_p-1) u_p^2 (\hat a \sqrt{u_p}-1)^2}{(1-6u_p+8\hat a u_p^{3/2}-3\hat a^2 u_p^2)(\hat a^2 u_p^2-2 u_p+1)(1-3 u_p+2\hat a u_p^{3/2})^{1/2}}+O(e^4)\,,
\end{eqnarray}
which correctly goes to zero far from the black hole where $u_p\to0$.

\end{widetext}

\subsection{Gyroscope moving along an unbound equatorial orbit}

In the case of unbound orbits not captured by the black hole, Eq.~\eqref{rdichi} represents a classical Newtonian parabolic ($e=1$) or hyperbolic ($e>1$) orbit which precesses due to general relativity,  with a minimal radius $r_{\rm per}$ of closest approach.  We consider only the hyperbolic-like orbits of the latter type resembling a
a classical scattering  process in which the geodesic path does not circle the black hole more than once, which occurs as long as the periastron is not too close to the black hole. Mathematically this corresponds to $\phi(\chi_{\rm (max)})<\pi$.
We compare the direction of the spin of the gyroscope before starting its gravitational interaction with the black hole (i.e., at $\tau \to -\infty$) with that after their interaction (i.e., at $\tau \to \infty$).
The relativistic anomaly now varies in the range $\chi\in[-\chi_{\rm (max)},\chi_{\rm (max)}]$, where $\chi_{\rm (max)}={\rm arccos}( -{1}/{e})$. 

For computational purposes it is convenient to parametrize the orbit instead in terms of the dimensionless inverse radial variable $u=M/r$, such that  
\begin{eqnarray}
\left(\frac{du}{d\tau}\right)^2&=&\frac{2\hat x^2}{M^2}u^4(u-u_1)(u-u_2)(u-u_3)\,,\nonumber\\
\frac{d\phi}{d\tau}&=&\frac{2\hat x}{M\hat a^2}u^2\frac{u_4-u}{(u-u_+)(u-u_-)}\,.
\end{eqnarray}
Here $u_1<u_2<u_3$ are the ordered roots of the equation
\beq
u^3-(\hat x^2+2\hat a\hat E\hat x+\hat a^2)\frac{u^2}{2\hat x^2}+\frac{u}{\hat x^2}+\frac{\hat E^2-1}{2\hat x^2}=0\,,
\eeq
whereas
\beq
u_\pm=\frac{M}{r_\pm}\,,\quad
u_4=\frac{L}{2x}\,.
\eeq
For hyperbolic orbits $u_1<0<u\leq u_2<u_3$, with $u_2$ corresponding to the distance of closest approach.

The rate of gyroscope precession and azimuthal change are given by
\begin{eqnarray}
\label{dudphi_geo}
\frac{d\phi}{du}&=&\pm\frac{\sqrt{2}}{\hat a^2}\frac{u_4-u}{(u-u_+)(u-u_-)}\times \nonumber\\
&& \frac1{\sqrt{(u-u_1)(u_2-u)(u_3-u)}}\,,\nonumber\\
\frac{d\Psi}{du}&=&\pm\frac{\hat a +E\hat x}{\sqrt{2}\hat x(1+\hat x^2u^2)}\times \nonumber\\
&& \frac1{\sqrt{(u-u_1)(u_2-u)(u_3-u)}}\,,
\end{eqnarray}
which can be integrated in terms of elliptic functions. 
Since the scattering process is symmetric with respect to the closest approach distance, the total change results from twice the integration  between $0$ and $u_2$, i.e., 
\beq
\Phi=2\int^{u_2}_0\frac{d\phi}{du}du\,,\qquad
\Psi=2\int^{u_2}_0\frac{d\Psi}{du}du\,,
\eeq
where we have assumed $\Phi(u_2)=0=\Psi(u_2)$ and the plus sign must be selected in Eq.~\eqref{dudphi_geo}.
The following explicit expressions in terms of elliptic functions hold

\begin{widetext}

\begin{eqnarray}
\Phi&=&-\frac{4\sqrt{2}}{\hat a^2(u_+-u_-)\sqrt{u_3-u_1}}\left[
\frac{u_4-u_+}{u_1-u_+}\left(\Pi(\alpha,\beta_+,m)-\Pi(\beta_+,m)\right)
-\frac{u_4-u_-}{u_1-u_-}\left(\Pi(\alpha,\beta_-,m)-\Pi(\beta_-,m)\right)
\right]
\,,\nonumber\\
\Psi&=&-i\frac{\sqrt{2}m(\hat a+E\hat x)}{\hat x^2(u_2-u_1)^{3/2}}\left[
\beta\left(\Pi(\alpha,\beta,m)-\Pi(\beta,m)\right)
-\bar\beta\left(\Pi(\alpha,\bar\beta,m)-\Pi(\bar\beta,m)\right)
\right]
\,,
\end{eqnarray}
where
\beq
m=\sqrt{\frac{u_2-u_1}{u_3-u_1}}\,,\quad
\alpha=\sqrt{\frac{-u_1}{u_2-u_1}}\,,\quad
\beta_\pm=\frac{u_2-u_1}{u_\pm-u_1}\,, \quad
\beta=\frac{i\hat x(u_2-u_1)}{1-i\hat xu_1}\,,
\eeq

\end{widetext}

and 
\beq
\Pi(\varphi,n,k)=\int_0^{\varphi}\frac{dz}{(1-n\sin^2z)\sqrt{1-k^2\sin^2z}}\,,
\eeq
with $\Pi(\pi/2,n,k)=\Pi(n,k)$ is the incomplete elliptic integral of the third kind~\cite{Gradshteyn}.

For the case of ``simple" scattering orbits under consideration here the total change in the azimuthal angle is less than $2\pi$ (less than a single revolution about the black hole). The scattering angle of the whole process can then be defined as
\beq
\frac12 \chi_{\rm scat}=\frac12 \Phi - \frac{\pi}{2}\,,
\eeq
with $\chi_{\rm scat}<\pi$.

To compare with existing literature (see e.g., Ref. \cite{Bini:2017wfr}) we define an energy-related variable $\bar E$ 
\beq
\bar E= \frac12 \sqrt{E^2-1}
\eeq
in place of the energy per unit mass $E$ and a combined (dimensionless) variable
\beq
\alpha =\frac{1}{ \sqrt{2 \bar E j^2}} \,,
\eeq
defined through $\bar E$ and the (dimensionless) angular momentum per unit mass $j\equiv\hat L=L/M$ and used, in turn, in place of $\bar E$. 
Once all factors of $c$ are restored,
\beq
\bar E \to \bar c^2\,,\qquad j\to \frac{j}{c}\,,
\eeq
one can perform the PN-expansion of Eqs.~\eqref{dudphi_geo} and integrate order by order by taking the Hadamard's {\it partie finie}, following the prescriptions of Ref. \cite{Bini:2017wfr} (see Section III B there).
The final result for $\Psi=\Psi_0+\hat a\Psi_{\hat a}+\hat a^2\Psi_{\hat a^2} +O(\hat a^3)$ then has the form of a power series in $1/j$, that is
\begin{widetext}
\begin{eqnarray}
\frac12 \Psi_0 &=& B(\alpha)+\frac{1}{j^2}\left[\frac32 B(\alpha)+\frac12 \frac{(1+3\alpha^2)}{\alpha (\alpha^2+1)}\right]+\frac{1}{j^4}\left[\frac{3(2+35\alpha^2)}{8\alpha^2}  B(\alpha)+\frac{(-1+67\alpha^2+181\alpha^4+105\alpha^6)}{8 \alpha^3 (\alpha^2+1)^2}\right]\nonumber\\
&+& \frac{1}{j^6}\left[
\frac{3(-1+140 \alpha^2+770\alpha^4)}{16\alpha^4}B(\alpha)+\frac{3+193\alpha^2+5913\alpha^4+18597\alpha^6+19740\alpha^8+6930\alpha^{10}}{48\alpha^5(\alpha^2+1)^3}
\right]\nonumber\\
&+& O\left(1/j^8\right)\,,\nonumber\\
\frac12  \Psi_{\hat a} &=&-\frac{1}{j^3}\left[3 B(\alpha)+\frac{(1+3\alpha^2)}{\alpha (\alpha^2+1) }  \right]  -\frac{1}{j^5}\left[ \frac{3 (3+35\alpha^2)}{2\alpha^2 }B(\alpha) +\frac{ (71+184\alpha^2+105\alpha^4)}{2\alpha(\alpha^2+1)^2} 
 \right]\nonumber\\
&+&O\left(1/j^7\right)\,,\nonumber\\
\frac12  \Psi_{\hat a^2} &=&  \frac{1}{j^4}\left[ \frac{3}{2} B(\alpha)+\frac{(2+3\alpha^2)}{2\alpha (\alpha^2+1)}  \right] +\frac{1}{j^6}\left[
\frac{9(4+35\alpha^2)}{4\alpha^2}B(\alpha) +\frac{2+228\alpha^2+561\alpha^4+315\alpha^6}{4\alpha^3(\alpha^2+1)^2}
\right]\nonumber\\
&+& O\left(1/j^8\right)\,,
\end{eqnarray}
where
\beq
B(\alpha) ={\rm arctan}(\alpha)+\frac{\pi}{2}\,.
\eeq
Similarly
\begin{eqnarray}
\frac12  \Phi_0 &=& B(\alpha)+\frac{1}{j^2}\left[3  B(\alpha) +\frac{  (2+3\alpha^2)}{ (\alpha^2+1)\alpha }\right] +\frac{1}{j^4}\left[ \frac{15(1+7\alpha^2)}{4\alpha^2} B(\alpha)+\frac{ 81+190\alpha^2+105\alpha^4}{4\alpha (\alpha^2+1)^2} \right]\nonumber\\
&+& \frac{1}{j^6}\left[ \frac{105(3+11\alpha^2)}{4\alpha^2}B(\alpha)  +\frac{256+3663\alpha^2+10143\alpha^4+10185\alpha^6+3465\alpha^8 }{12\alpha^3(\alpha^2+1)^3} \right]\nonumber\\
&+& O\left(1/j^8\right)\,,\nonumber\\
\frac12  \Phi_{\hat a} &=& -\frac{1}{j^3}\left[ 4 B(\alpha)+2 \frac{(1+2 \alpha^2)}{ (\alpha^2+1) \alpha} \right]- \frac{1}{j^5}\left[  12\frac{(1+7\alpha^2)}{\alpha^2} B(\alpha)+\frac{(1+65\alpha^2+152\alpha^4+84\alpha^6)}{\alpha^3 (\alpha^2+1)^2} \right] \nonumber\\
&+& O\left(1/j^7\right)\,,\nonumber
\end{eqnarray}
\begin{eqnarray}
\frac12 \Phi_{\hat a^2} &=&\frac{1}{j^4}\left[  \frac32  B(\alpha)+\frac{(2+3\alpha^2)}{2(\alpha^2+1)\alpha}   \right] + \frac{1}{j^6}\left[
\frac{3(11+70\alpha^2)}{2\alpha^2}B(\alpha)+\frac{4+167\alpha^2+383\alpha^4+210\alpha^6}{2\alpha^3(\alpha^2+1)^2}
\right]\nonumber\\
&+& O\left(1/j^8\right)\,.
\end{eqnarray}
The above expression for $\Phi_0$ agrees with Eqs.~(45)--(46) (in the point-particle limit  $\nu=0$) of Ref. \cite{Bini:2017wfr}.

\end{widetext}

\section{Concluding remarks}

All the relevant observer-adapted frames needed to construct geometrically-motivated frames along a general timelike geodesic in a Kerr black hole spacetime are described in terms of combinations of boost operations applied to the natural orthonormal frames associated with Boyer-Lindquist coordinates. Thus Marck's seemingly arbitrary recipe for obtaining a parallel transported frame along timelike geodesics acquires a nice interpretation in terms of identifying a parallel transported axis of rotation and an angular velocity of rotation in the 2-plane orthogonal to it, anchored radially to the usual coordinate grid of the spacetime.
The sequence of boosts and rotations needed to interpret Marck's frame has now been made explicit.

Finally, we have discussed the cumulative precession effects of a test gyroscope moving along both bound and unbound equatorial plane geodesic orbits by evaluating the total spin-precession angle and the cumulative azimuthal phase as well as the associated spin precession invariant.

\appendix

\section{Boost and rotation matrices}

The 6-dimensional Lie algebra of the Lorentz matrix group is generated by $4\times4$ matrices corresponding to the components of mixed second rank tensors which are antisymmetric when the flat Minkowski metric (component matrix $(\eta_{\alpha\beta})={\rm diag}(-1,1,1,1)$) is used to lower or raise indices to a completely covariant or covariant form. Six linearly independent matrices from this set
\beq
  L_{\alpha\beta} = ([L_{\alpha\beta}]{}^\gamma{}_\delta)\,,\quad
[L_{\alpha\beta}]^{\gamma\delta}=\delta^{\gamma\delta}_{\ \alpha\beta}
  =2 \delta^{[\gamma}{}_{\alpha} \delta^{\delta]}{}_{\beta}
\eeq
form a basis of the matrix Lie algebra.
Three rotation generators are
\begin{widetext}
\beq
  J_1=(L_{23}{}^\alpha{}_\beta )
=\begin{pmatrix}
0 & 0 & 0 & 0 \cr
0 & 0 & 0 & 0 \cr
0 & 0 & 0 & -1 \cr
0 & 0 & 1 & 0
\end{pmatrix}
\,,\quad
J_2 = (L_{31}{}^\alpha{}_\beta )
=\begin{pmatrix}
0 & 0 & 0 & 0 \cr
0 & 0 & 0 & 1 \cr
0 & 0 & 0 & 0 \cr
0 & -1 & 0 & 0
\end{pmatrix}
\,,\quad 
J_3=(L_{12}{}^\alpha{}_\beta )
=\begin{pmatrix}
0 & 0 & 0 & 0 \cr
0 & 0 & -1 & 0 \cr
0 & 1 & 0 & 0 \cr
0 & 0 & 0 & 0
\end{pmatrix}
\eeq
and three boost generators are
\beq
 K_1=(L_{01}{}^\alpha{}_\beta )
=\begin{pmatrix}
0 & 1 & 0 & 0 \cr
1 & 0 & 0 & 0 \cr
0 & 0 & 0 & 0 \cr
0 & 0 & 0 & 0
\end{pmatrix}
\,,\quad
K_2 = (L_{02}{}^\alpha{}_\beta )
=\begin{pmatrix}
0 & 0 & 1 & 0 \cr
0 & 0 & 0 & 0 \cr
1 & 0 & 0 & 0 \cr
0 & 0 & 0 & 0
\end{pmatrix}
\,,\quad 
K_3=(L_{03}{}^\alpha{}_\beta )
=\begin{pmatrix}
0 & 0 & 0 & 1 \cr
0 & 0 & 0 & 0 \cr
0 & 0 & 0 & 0 \cr
1 & 0 & 0 & 0
\end{pmatrix}
\eeq
\end{widetext}

The rotation and boost matrices  are obtained exponentiating linear combinations of these two subsets of the Lie algebra, collapsing the exponential series to three terms from the cubic identities
\beq
  (n^i J_i)^3 = -n^i J_i\,,\quad
  (n^i K_i)^3 = n^i K_i
\eeq
satisfied for a unit vector $\delta_{ij}n^i n^j=1$. One finds
\begin{eqnarray}\label{rotationboostmatrices}
  R(\varphi,n^i) 
&=& e^{\varphi\, n^i J_i}\nonumber\\
&=& Id + \sin\varphi\, (n^i J_i) -(\cos\varphi-1)\, ( n^i J_i)^2\,,
\nonumber\\
  B(\alpha,n^i) 
&=& e^{\alpha\, n^i K_i}\nonumber\\
&=& Id + \sinh\alpha\, (n^i K_i) +(\cosh\alpha-1)\, ( n^i K_i)^2\,.\nonumber\\
\end{eqnarray}
These represent an active rotation by an angle $\theta$ of the 2-plane orthogonal to $n^i$ in space (in the direction related to $n^i$ by the right hand rule), and an active boost by the rapidity $\alpha$ along the spatial velocity $\nu^i=\tanh\alpha\, n^i$. The boost may instead be parametrized by the relative velocity components $\nu^i$ themselves
\beq
  \mathcal{B}(\pmb{\nu}) \equiv \mathcal{B}(\nu^i)=B(\alpha,n^i)
\,.
\eeq

The black hole equatorial plane case corresponds to the 3-dimensional Lorentz subgroup generated by $J_3,K_1,K_2$, with boost matrices $\mathcal{B}(\nu^1,0,\nu^3)$ and rotations $ R(\varphi,n^1,0,n^3) $.

\end{document}